\documentclass[pre,preprint]{revtex4}
\usepackage{amsmath}
\usepackage{amsfonts}
\usepackage{graphicx}

\begin{document}

\title{The effect of disorder in the contact probability of elongated conformations of biopolymers}
\author{Guido Tiana}
\affiliation{Department of Physics, Universit\`a degli Studi di Milano and INFN, via Celoria 16, 20133 Milano, Italy}

\begin{abstract}
Biopolymers are characterized by heterogeneous interactions, and usually perform their biological tasks forming contacts within domains of limited size. Combining polymer theory with a replica approach, we study the scaling properties of the probability of contact formation in random heteropolymers as a function of their linear distance. It is found that close or above the theta--point, it is possible to define a contact probability which is typical (i.e. "self-averaging") for different realizations of the heterogeneous interactions, and which displays an exponential cut--off, dependent on temperature and on the interaction range. In many cases this cut--off is comparable with the typical sizes of domains in biopolymers. While it is well known that disorder causes interesting effects at low temperature, the behavior elucidated in the present study is an example of a non--trivial effect at high temperature.
\end{abstract}

\date{\today}
\maketitle

In biopolymers, the formation of contacts between monomers in non--compact conformations is one of the basic physical procceses which eventually determine the function of the molecule. For example, in the case of proteins, the formation of non--covalent interactions between distant amino acids in the denatured state is, in many cases, among the first steps in the folding process \cite{Bruun:2010}. In fact, the folding rate of proteins from their denatured state has been shown to be correlated with the  separation  $\Delta l$ along the chain of the pairs of residues which are in contact in the native conformation \cite{Plaxco:1998}, and the same phenomenon was observed in minimal protein models \cite{Cieplak:1999}. Similarly, in chromatin the contact probability between loci was found to correlate with their linear distance on the megabase scale \cite{LiebermanAiden:2009jz,Naumova:2013fg}.

Moreover,  biopolymers usually display a rather tight upper limit in the value of $\Delta l$ associated with their contacts. Most of them are structured in domains of characteristic size, and the formation of contacts takes place predominantly within such domains. For example, proteins can be very long, but the distribution of domain sizes drops above 250 residues \cite{Xu:1997}, while longer proteins are usually built of multiple domains that fold independently and then assembly together. Also in the case of chromatin, the polymer seems to form domains with a maximum size of the order of $10^5-10^6$ bases \cite{Sexton:2015in}.

The simplest description we can give to the contact formation in biopolymer is through a homopolymeric model at equilibrium. In the elongated states of homopolymers, the contact probability between two monomers depends on  $\Delta l$ (when this is sufficiently large) according to a power law $\Delta l^{-\kappa}$, where $\kappa=3/2$ for ideal chains, that is when the effective interaction between monomers is null,  and $\kappa=9/5$ \cite{deGennes} for a random coil made of mutually repuslive monomers. As a matter of fact, the contact probability between the ends of unstructured peptides with repeated AGQ sequence of length up to 29 residues, measured by FRET, displays a power--law with respect to the length of the peptide whose exponent changes from 1.55 in water to 1.7 in urea and guanidine \cite{Buscaglia:2006}.  Simulations of the unfolded state of globular, single--domain proteins up to lengths of 250 monomers show a power--law dependence of contact probabilities with exponent 2.0 \cite{Ding:2005}. Even the crystal structure of proteins seems remniscent of the associated denatured state and displays ideal--chain statistics \cite{Banavar:2005}, the distribution of loop sizes having a maximum at 27 residues \cite{Berezovsky:2000}. 

A homopolymeric model can account for the power--law relation between contact probability and linear distance, but  it cannot explain the presence of finite-size domains. On the contrary, the long tail associated with power--laws would suggest that the probability of  non--local contacts remains rather high even at large separation distances. In such a scenario, the evolutive advantage of shaping a biopolymer into domains to reduce the entropic cost of forming the initial contacts is weak or null.

However, biomolecules are rarely homopolymers, and consequently it may be useful to investigate the effect of the heterogeneity of the interactions in the probability of contact formation. As a model, we will consider a random heteropolymer \cite{Shakhnovich:1988wi,Shakhnovich:1989tm} interacting with the potential
\begin{equation}
U(\{r_i\})=U_0(\{r_i\}) + \frac{v}{2}\sum_{ij}^N B_{ij} \delta(r_i-r_j),
\label{eq:u}
\end{equation} 
where $U_0(\{r_i\})=\sum_i  u_0(|r_i-r_{i-1}|-a)$ is any function that mantains the integrity of the polymer, $a$ is the inter--monomer separation length, $N$ the number of monomers, $v$ is the interaction volume and $B_{ij}$ accounts for two-body interactions, assumed as stochastic quenched variables distributed according to a Gaussian of average $B_0$ and standard deviation $\sigma$.

One would be interested in the contact probability 
\begin{equation}
R_{ij}\equiv\frac{Z_{ij}}{Z}\equiv\frac{   v\int d\{r\} \,\delta(r_i-r_j) \exp[-\beta U(\{r\})]   }{ \int d\{r\} \exp[-\beta U(\{r\})]  }
\end{equation}
 between monomers $i$ and $j$ of the chain. But in disordered systems, relevant quantities should be averaged over disorder, and this is meaningful only is if their relative fluctuations become negligible when the system is large enough, namely if they are self--averaging \cite{Lifshitz:1942}. The standard Brout's argument \cite{Brout:1959} suggests that extensive quantities, like the free energy, are self--averaging. The argument says that in a system with a given realization of the disordered interactions, the relative fluctuations of extensive quantities go to zero in the thermodynamic limit thanks to the central limit theorem. If one divides this system in $K$ weakly--interacting sub--systems, each of them can be regarded as a different realization of the disordered interaction in an identical, although smaller, system, and consequently the relative fluctuations of the extensive quantity over the disorder goes to zero as well. 
 
The quantity $F_{ij}=-T\log R_{ij}$ is a free energy, but it is difficult to apply Brout's argument to it. In fact, when dividing the whole chain into weakly--interacting sub--systems, these will not be identical to each other, because one of them will contain the loop and the others not. Instead, a self--averaging quantity is expected to be $F_{\Delta l}=-T(N-\Delta l)^{-1}\sum_m\log R_{m,m+\Delta l}$ in the case that $\Delta l\ll N$. In fact, dividing the chain in $K$ subsystems, $F_{\Delta l}$ results the sum of $(N-\Delta l)(K-1)$ terms accounting for the free energy of an unconstrained polymer, and $(N-\Delta l)$ terms accounting for the free energy of a looped polymer. The relative fluctuations of these sums, which can be regarded as averages over the disordered interactions, go to zero for $N-\Delta l\gg 1$, suggesting that $F_{\Delta l}$ is self--averaging.

The average of $F_{\Delta l}$ over disorder can be evaluated with the standard replica trick \cite{Mezard:book} from
\begin{equation}
\overline{\log R_{\Delta l}}\equiv\frac{1}{N-\Delta l}\sum_m\overline{\log R_{m,m+\Delta l}}=\frac{1}{N-\Delta l}\sum_m\lim_{n\to 0}\frac{1}{n}\log  \frac{ \overline{Z^n_{m,m+\Delta l}} }{ \overline{Z^n}  }.
\label{eq:logr}
\end{equation}
The  constrained partition function $Z^n_{ij}$ defined by a contact $i-j$ can be integrated over the Gaussian--distributed interaction elements $B_{ij}$ as in ref. \cite{Sfa:1993} to give
\begin{align}
&\overline{Z^n_{ij}} = \nonumber\\ &=\int d\{r^\alpha\} \left(v^{n}\prod_{\alpha}\delta(r_i^\alpha-r_j^\alpha)\right)   \exp\left[-\beta \sum_\alpha U_1(\{r^\alpha_l\}) + \frac{\beta^2\sigma^2}{2} v^{2}\sum_{k\neq l \, \alpha\neq\beta} \delta(r_k^\alpha-r_l^\alpha) \delta(r_k^\beta-r_l^\beta) \right],
\label{eq:zn1}
\end{align}
where $U_1= \sum_k u_0(|r_k^\alpha-r_{k-1}^\alpha|-a)+\frac{vB'_0}{2} \sum_{kl\alpha}\delta(r_k^\alpha-r_l^\alpha)$ with $B'_0=B_0-\beta\sigma^2$ is the effective one--replica interaction which controls the density of the chain \cite{Shakhnovich:1989wr}. A similar expression, lacking of the product of $\delta(r_i^\alpha-r_j^\alpha)$ holds for the unconstrained $\overline{Z^n}$.

For each pair of replicas, the double sum over monomers at exponential of Eq. (\ref{eq:zn1}) counts the number of contacts shared by the two replicas. If the chain has density $\rho$, this number is expected to scale as $N\rho^2$, because each monomer has a probability $\rho$ to be in contact with another monomer of the same replica, and a probability $\rho$ to be in contact with the same monomer within the other replica. Thus, the number of shared contact results independent on $N$ close to the $\theta$--point, where $\rho=\rho_0\sim N^{-1/2}$, and decreases above the $\theta$--point, where $\rho<\rho_0$. If $v\ll a^3$, for "biological temperatures" ($\beta^{-1}\sim\sigma$)  the term which couples the replica together can be treated perturbatively, obtaining
\begin{align}
\overline{Z^n_{ij}}  &=  \int d\{r^\alpha\} \left(v^{n}\prod_{\alpha}\delta(r_i^\alpha-r_j^\alpha)\right)\times \nonumber\\
& \times \exp\left[-\beta \sum_\alpha U_1(\{r^\alpha_l\}) \right] \left( 1+\frac{\beta^2\sigma^2}{2} v^{2}\sum_{k\neq l\,\alpha\neq\beta} \delta(r_k^\alpha-r_l^\alpha) \delta(r_k^\beta-r_l^\beta) \right).
\end{align}
At variance with the perturbation approach applied to the excluded--volume \cite{deGennes}, in which case the perturbing term scales as $(T-\theta)N^{1/2}$ and thus is meaningful only if $T\sim\theta$, in the present case it can be applied at any temperature.
Following a scheme similar to that of ref. \cite{Chan:1989un} one can write the ratio needed in Eq. (\ref{eq:logr}) as
\begin{equation}
\lim_{n\to 0}\frac{1}{n}\log  \frac{ \overline{Z^n_{ij}} }{ \overline{Z^n} } = \lim_{n\to 0}\frac{1}{n}\log \left(  \frac{ \overline{(Z^{(0)}_{ij})^n } }{ \overline{(Z^{(0)})^n} }
\left[ 1 + \left(  \frac{ \overline{(Z^{(2)}_{ij})^n } }{ \overline{(Z^{(0)}_{ij})^n} } - \frac{ \overline{(Z^{(2)})^n } }{ \overline{(Z^{(0)})^n} }  \right) \right] \right)
\label{eq:zsuz}
\end{equation}
where the superscript $(2)$ indicates the perturbed partition functions, including the term proportional to $v^2$, while the superscript $(0)$ indicates the unperturbed partition function. The last fraction is imamterial because it does not depend on $(j-i)$.

The perturbed constrained partition function can be written as 
\begin{equation}
\overline{(Z^{(2)}_{ij})^n}=v^{4}\beta^2\sigma^2\frac{  \overline{(Z^{(0)}_{ij})^n}   }{   \overline{(Z^{(0)}_{ij})^2} }  \; n(n-1)\sum_{k< l}  \left( \int dr  \delta(r_i-r_j)  \delta(r_k-r_l)    e^{-\beta U_1(r)} \right)^2
\label{eq:z2ij}
\end{equation}
where the integral is performed over a single replica. The above sum can be split in terms defined by the order of the indexes $k$, $l$, $i$ and $j$, which can be graphically represented as
\begin{equation}
 \sum_{k<l}
\left[  
\raisebox{-3ex}{\includegraphics[height=7ex]{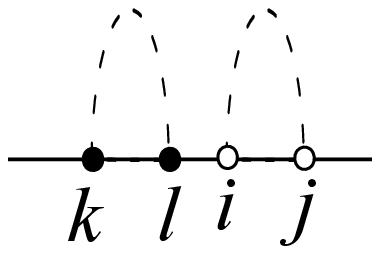}} +
\raisebox{-3ex}{\includegraphics[height=7ex]{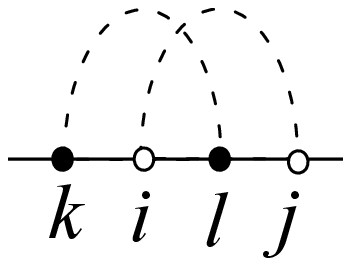}} +
\raisebox{-3ex}{\includegraphics[height=7ex]{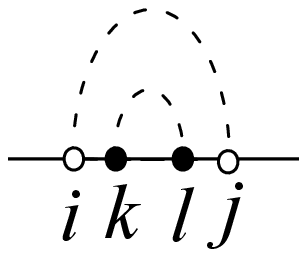}} +
\raisebox{-3ex}{\includegraphics[height=7ex]{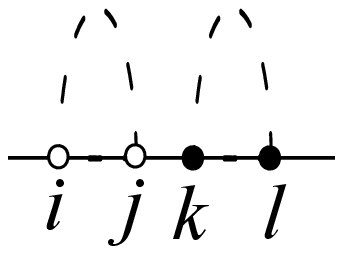}} +
\raisebox{-3ex}{\includegraphics[height=7ex]{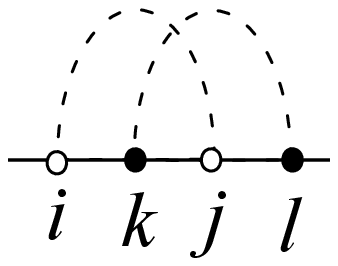}} +
\raisebox{-3ex}{\includegraphics[height=7ex]{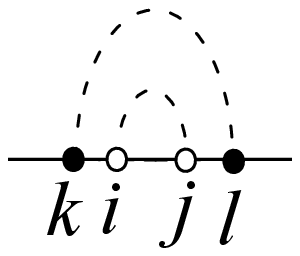}}
\right].
\label{eq:fey}
\end{equation}
To calculate the integrals in the sums defined by the above graphs, we employ the approximation that $U_1\approx U_0$, so that the propagators associated with the solid segments in the graphs is that of the ideal chain,
\begin{equation}
G_0(\Delta r,\Delta l)=\frac{1}{(2\pi a^2\Delta l)^{3/2}}\exp\left[-\frac{|\Delta r |^2}{2a^2\Delta l}  \right],
\end{equation}
which is exact if $B'_0=0$ and worsen as $B'_0>0$ increases. Vertexes are $V(\Delta r)=2\pi\delta(|\Delta r|)$. Each term of Eq. (\ref{eq:fey}) can be calculated integrating the chain of propagators corresponding to the graph and approximating the sum over $k$ and $l$ as  integrals. For example, the simplest contribution is
\begin{align}
\sum_{k<l}\raisebox{-3ex}{\includegraphics[height=7ex]{feynman1.eps}} &=\sum_{k<l}^i\left(\Omega\int d^3r_kd^3r_ld^3r_id^3r_jd^3r_N G_0(r_k,k)G_0(r_l-r_k,l-k)V(r_l-r_k)\times\right.\nonumber\\
&\left.\vphantom{\int} \times G_0(r_i-r_l,i-l)G_0(r_j-r_i,j-i)V(r_j-r_i)G_0(r_N-r_j,N-j) \right)^2 =\nonumber\\
&=\frac{\Omega^2(i-2)(i^2-2i+2)}{8\pi^2a^{12}(i-1)^2}\cdot\frac{1}{(j-i)^3},
\label{eq:term1}
\end{align}
where $\Omega$ is the volume of conformational space of the chain, needed because the functions $G_0$ are probability densities. In four of the six terms of Eq. (\ref{eq:fey}), the leading contribution in the limit of large $(j-i)$ scales as   $(j-i)^{-3}$. Exception is made for the third and the sixth term, which give
\begin{align}
\sum_{k<l}\raisebox{-3ex}{\includegraphics[height=7ex]{feynman3.eps}} &=\frac{\Omega^2}{8\pi^2a^{12}}\cdot \frac{1}{(j-i)^2} \nonumber \\
\sum_{k<l}\raisebox{-3ex}{\includegraphics[height=7ex]{feynman6.eps}} &=\frac{\Omega^2}{8\pi^2a^{12}(\min[j,N-i+1])^2}\cdot \frac{1}{(j-i)^2}
\label{eq:terms2}
\end{align}
These two terms give a perturbation to $\overline{\log R_{ij}}$ which depends on $(j-i)$, since the square of the unperturbed constrained partition function with which one must compare them (see Eqs. (\ref{eq:zsuz}) (\ref{eq:z2ij}), and below) scale as $(j-i)^{-3}$ for the ideal chain and $(j-i)^{-18/5}$ for the swollen coil.

\begin{figure}
\includegraphics[width=\textwidth]{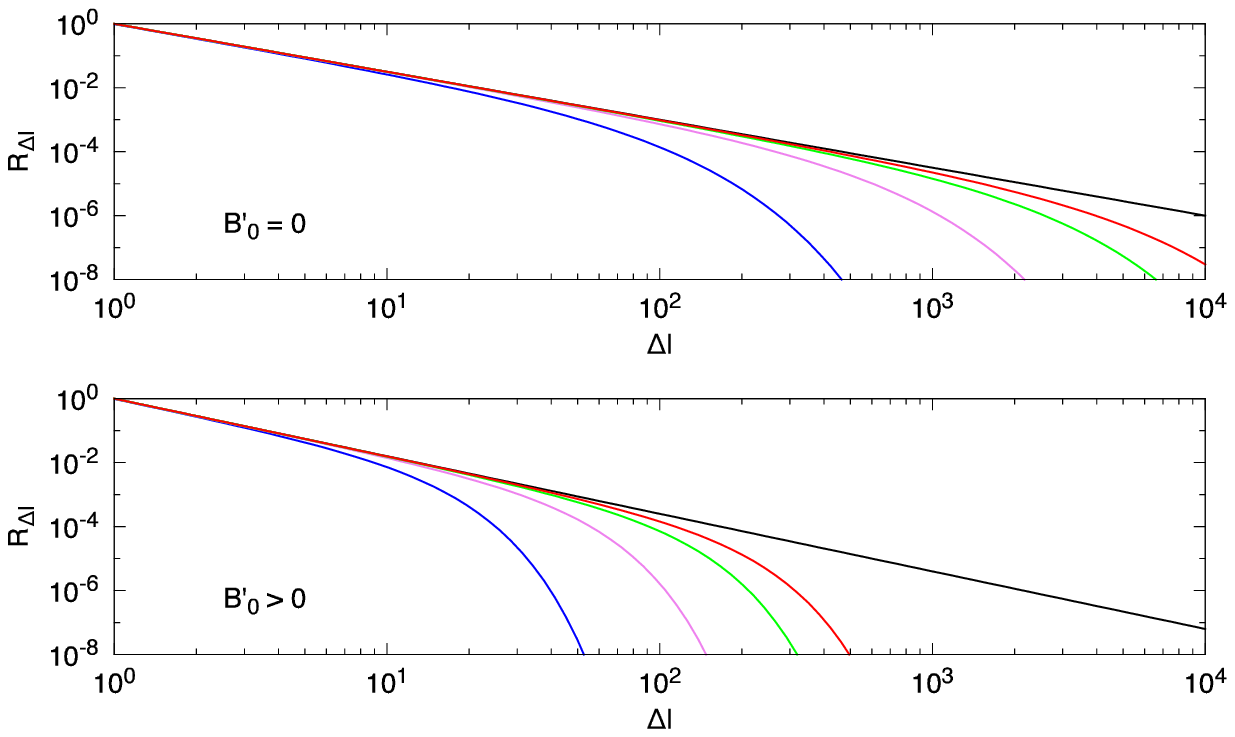}
\caption{The shape of $R_{\Delta l}$ calculated at $T=0.2\sigma$ (blue curve), $T=0.5\sigma$ (purple curve), $T=\sigma$ (green curve) and $T=1.5\sigma$ (red curve), for the cases $B'_0=0$ (upper panel) and  $B'_0>0$ (lower panel). The black curves show the power law without exponential correction.}
\label{fig1}
\end{figure}

Identifying the conformational--space volume $\Omega$ with $\overline{(Z^{(0)})^n}$, Eq. (\ref{eq:logr}) can be calculated using Eqs. (\ref{eq:zsuz}), (\ref{eq:z2ij}) and (\ref{eq:terms2}), and performing the limit $n\to 0$, resulting for large $\Delta l$ in
\begin{equation}
\overline{\log R_{\Delta l}}=\log\left(  \frac{1}{\Delta l^{\kappa}} \cdot \exp\left[-\left( \frac{ \Delta l }{\Delta l_0} \right)^{2\kappa-2}  \right]   \right),
\label{eq:summa}
\end{equation}
where
\begin{equation}
\Delta l_0\equiv \left[\frac{8\pi^2}{\beta^2\sigma^2}\left( \frac{a^3}{v}  \right)^4\right]^{\frac{1}{2\kappa-2}}.
\label{eq:dl0}
\end{equation}
This means that when one plots the contact probabilities of pairs of monomers versus their separation in log--log scale, the linear behaviour can be detected only for $\Delta l\ll \Delta l_0$, while an exponential drop dominates at larger values of $\Delta l$. The exponent associated with the power--law regime does not result to change with respect to the homopolymeric case, as already suggested in ref. \cite{Stepanow:1992}  making use of the renormalization group in three dimension. Examples of the exponential correction to the power law are displayed at different temperatures  in Fig. \ref{fig1} for the case of the ideal chain (upper panel) and the random coil (lower panel).

When the temperature is close to the $\theta$--point (i.e., $B'_0=0$), then $\kappa=3/2$ and the effect of the disorder in the interactions is that of applying an exponential cutoff to the contact probability to monomers whose linear distance is beyond $\Delta l_0$. To have an order of magnitude for  $\Delta l_0$ in the ideal--chain regime, one can consider the case $B_0=\sigma$, so that $B'_0=0$ implies $\beta^2\sigma^2=1$ (i.e., $T=\sigma$), and $v/a^3=0.5$. In this case,  $\Delta l_0\approx 10^3$.

Above the theta point  (i.e., $B'_0>0$) $\kappa=9/5$ and the correction to the power--law behaviour is a stretched exponential with power $8/5$, which is not very far from a Gaussian function. If, for example, we still set $B_0=\sigma$ and $v/a^3=0.5$, but now choose $T=3/2\,\sigma$, so that $B'_0=1/3\,\sigma>0$ and the chain is in a coil phase, now $\Delta l_0\approx 130$. The coil regime, above the theta point, is the typical case experienced by proteins at the beginning of the folding process in the experiments \cite{Haran:2012}.

\begin{figure}
\includegraphics[width=\textwidth]{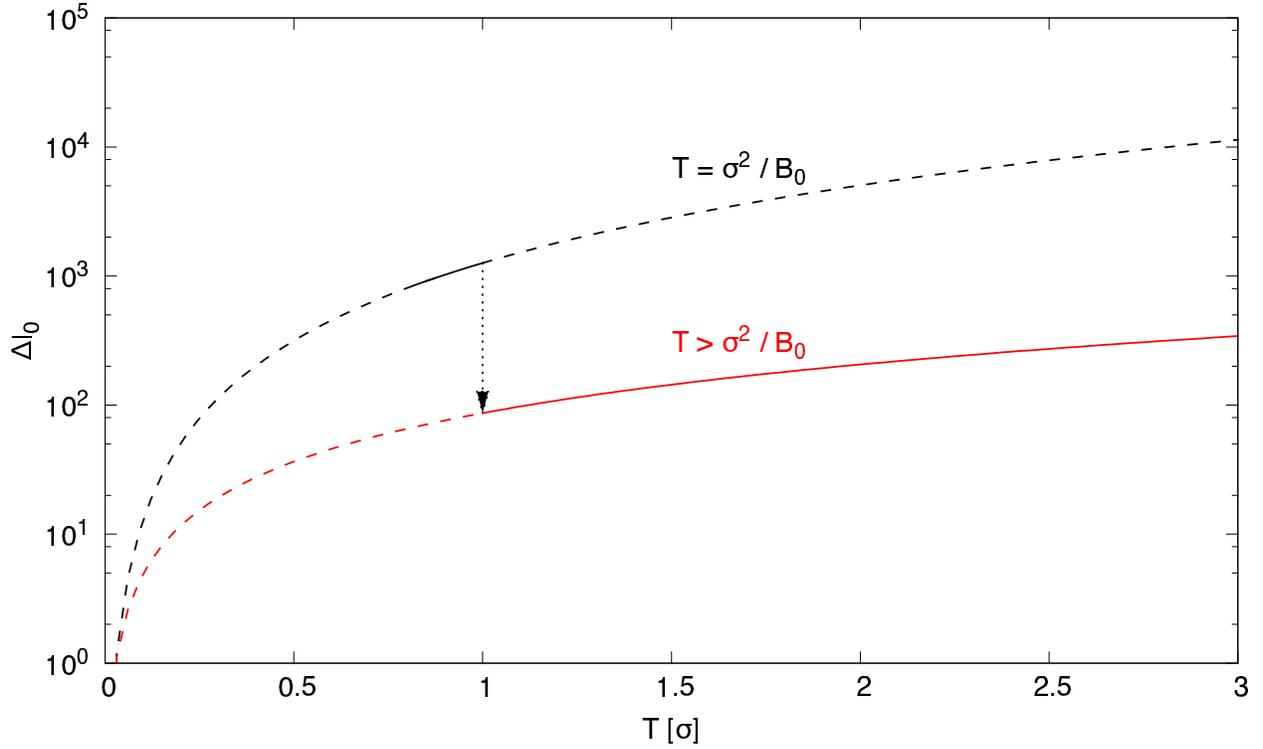}
\caption{The dependence of $\Delta l_0$ on the temperature for the cases $B'_0=0$ (black curve) and  $B'_0>0$ (red curve). If one assumes, for instance, that $B_0=\sigma$, then the transition between the two cases occurs at $T=\sigma$ (marked by a dotted arrow). At lower temperatures the system is in a globular phase and the presente calculations do not apply.}
\label{fig2}
\end{figure}

The behaviour of $\Delta l_0$ with respect to the temperature is quite irregular (see Fig. \ref{fig2}). At the theta point it can be large because, in spite of the small denominator in Eq. (\ref{eq:dl0}), the overall exponent is 1. When the temperature increases just above the theta point, the denominator becomes somewhat larger, but the overall exponent drops to 5/8, making $\Delta l_0$ small. As temperature is further increased, $\Delta l_0$ becomes larger and eventually diverges. A consequence of this is that, for each value of $B_0$, there is an intermediate range of temperatures which penalizes the formation of long--range structure in the elongated conformations of biopolymers.

Actual biopolymers are finite systems. The upper limit of their length is usually well--defined, and long biopolymers are structured in domains compatible with this upper limit.
In the case of proteins, single domains are shorter than $\sim 250$ residues \cite{Xu:1997}, corresponding to $\sim 100$ Kuhn lenghts. Among the factors which constrain the size of single domains could be the difficulty of establishing long--range contacts in the denatured state due to the exponential cutoff highlighted above, and consequently of achieving an efficient folding mechanism. Also in the case of chromatin, whose Kuhn length is $\sim 6\cdot 10^3$ bases \cite{Dekker:2008}, the polymer seems to form domains with a maximum size of the order of $10^5-10^6$ bases \cite{Sexton:2015in}, corresponding to hundreds of Kuhn lenghts.

The low--temperture globular phases of random heteropolymers have been widely studied in the past \cite{Sfa:1993,Shakhnovich:1989wr}, and show glassy behaviour. Interestingly, the effect of disorder on the contact probability displayed by Eq. (\ref{eq:summa})  appears at rather high temperatures, well above the glassy transition.

\begin{acknowledgments}
The author would like to thank Eugene Shakhnovich for helpful discussions and suggestions.
\end{acknowledgments}

\newpage

\end{document}